\documentclass[twoside]{dis09}
\usepackage[latin1]{inputenc}
\usepackage[dvips]{graphicx,epsfig,color}
\usepackage{wrapfig,rotating}
\usepackage{amssymb,amsmath,array}

\begin{document}
\title{The Large Hadron Electron Collider Project}

\author{Max Klein$^1$
%
\thanks{On behalf of the LHeC Steering Group.}
%
\vspace{.3cm}\\
%
1- University of Liverpool, Dept. of Physics, L69 7ZE Liverpool,
Oxford Street, UK.
%
}

\maketitle

\begin{abstract}
A Conceptual Design Report (CDR) for the Large Hadron Electron Collider, the LHeC,
is being prepared to which an introduction was given~\cite{url} 
for the plenary panel discussion on the future
of deep inelastic scattering held at DIS09. This is briefly summarised here.
The CDR will comprise designs of
the ep/eA collider, based on ring and linear electron accelerators,
of the interaction region, designed for simultaneous $ep$ and $pp$ operation,
of a new, modular detector, and it will
present basics on the physics motivation for a high luminous Tera scale
electron-nucleon collider as a complement to the LHC. \end{abstract}
\section{Introduction}
Much of the current thinking on the future of particle physics is focused
on the mechanism of electroweak symmetry breaking. Current limits from
electroweak precision measurements and from the Tevatron experiments
restrict the mass range of the standard model Higgs particle mass to a 
narrow window above the direct LEP limit, and with increasing intensity
one considers alternative mechanisms for the origin of mass of the vector
bosons~\cite{quigg}. The origin of baryonic mass is as fascinating a problem.
It is not rooted in the electroweak symmetry breaking but
in the dynamics of parton interactions as described in QCD.
With the possibility of standard expectations not becoming confirmed and
experiment, with the LHC, moving into Terascale physics, the prospects for
new physics become wider and the need for precision measurements
in a new kinematic range apparent. A new electron-proton and electron-ion collider
operating at the Terascale certainly would be a major instrument to study the
new horizon of particle physics and contribute substantially to its further
development  based on high energy particle accelerators.

The LHeC has a fascinating physics programme,
as has been presented to this workshop~\cite{uta,olaf,nestor}
and will be summarised below.
Based on current considerations, also presented at this conference, both the
machine~\cite{holzer} and a new detector~\cite{alessandro}
represent as fascinating opportunities. In a broader context 
a TeV $ep$ collider is related to a number of fundamental questions
worth dealing with prior to a more detailed discussion of its physics and
its technical challenges. Apart from the strong force, there is not much
difference between leptons and quarks, and it still is a mystery why these
form two types of matter instead of one  as Abdus Salam had already  
observed three decades ago~\cite{salam}. HERA was mankind's highest 
resolution microscope and established a limit of $6 \cdot 10^{-19}$\,m for
quarks to be pointlike.  Nowadays much effort is devoted
to build microscopes to study molecules using particle accelerators, or to look to the
outer space.   A new, high resolution microscope is due to move the frontier
of exploring the substructure of matter to even smaller
dimensions. Here, resonant electron-quark
states or a further layer of substructure may be found, at distances smaller
than four orders of magnitude than the radius of the proton. The LHeC
probes a quark radius as small as $7 \cdot 10^{-20}$\,m.  If a proliferation
of new states, possibly obeying supersymmetry, would be discovered at the LHC,
it was for an $eq$ collider based on lepton charge and polarisation variations
to unfold much of the new spectroscopy and determine the quantum numbers of
new states. If physics signals from still higher scales should occur at
the LHC, as via contact interaction enhancements, a Terascale
$ep$ collider was necessary to distinguish signs for new physics from
variations of parton distributions by virtue of high precision measurements
of partons into the kinematic range of the LHC.  
At the LHeC radically new phenomena in the dynamics 
of parton interactions are expected to be established, as  the saturation 
of the rise of the gluon density or instantons, and the mechanism of parton
emission at low $x$, often related to the BFKL or BK equations, may
be clarified.  For the first time ever,
owing to the high energy, luminosity and variation of charge and polarisation,
a DIS experiment will be able,
completely and essentially directly, to unfold the partonic content of the proton by
measuring the valence, light and all heavier sea quark and the gluon distributions.
The LHeC would extend the knowledge on the structure of the neutron and of nuclei
by four orders of magnitude. Thus it represented the necessary base for the study
of deconfinement and collective phenomena in nuclei, and it would support
the development of postulated links between QCD and string theory.

The present write-up describes the brief introduction~\cite{url} given on the LHeC 
project~\cite{lhecweb}
to the panel discussion on the future of deep inelastic scattering at the
DIS09 workshop.  The panel  also
discussed the options  for lower energy electron-nucleon and polarised $ep$ collisions
as are considered at BNL and Jlab~\cite{abhay}.
Previous summaries of the status of the LHeC project have been
given to DIS08~\cite{maxdis08}, with a basic discussion on the ring-ring (RR)
and the linac-ring (LR) configurations, and to ECFA in 2008~\cite{lhecweb}.

The intense proton and ion beams of the LHC
are the basis for a new $ep$ and $eA$ collider of unprecedented luminosity,
$100$ times larger than achieved at HERA. Thus CERN, ECFA and NuPECC have joined
efforts in designing a reliable and affordable concept.
The CDR on the LHeC is being worked out within a joint workshop,
see~\cite{lhecweb}, the first of which took place in 2008 with a follow-up meeting
in 2009. The CDR is foreseen to be delivered in 2010, further design work
is being pursued on the ILC, CLIC and a multi-TeV muon 
collider,  and with the
first observations at the LHC expected soon, the deliberations on the future
of HEP will eventually rely on solid experimental grounds.
A strategic goal may be to explore the Terascale phenomena with
complementary $pp$, $l^+l^-$ and $ep$ colliders, guided
by future expectations and in view of the successful  exploration
of the Fermi scale  with the Tevatron,
LEP and HERA experiments, now being completed.
\section{The Accelerator Alternative}
As is discussed e.g. in~\cite{maxdis08,fpac09}, two options for the 
electron beam are under
study, one based on a synchrotron and one using an electron linac.
An electron ring with newly designed magnets 
installed above the LHC  would provide 
energies between a few tens of GeV up to about $80$ GeV, as
is limited by synchrotron radiation losses and rf. A first rather
complete design study is~\cite{jinst}.
At $E_e=50$\,GeV and $P=50$\,MW power, such a configuration may deliver
$5 \cdot 10^{33}$\,cm$^{-2}$s$^{-1}$ luminosity, i.e.
a factor of $100$ higher than the maximum achieved 
with HERA.  Such values may be obtained in 
the standard LHC beam configuration and slightly
increase with the  sLHC upgrade~\cite{holzer}.
Such a luminosity potential may provide integrated
luminosities of order 100\,fb$^{-1}$. 

An electron linac may be independently installed providing 
energies between a few tens of GeV up to perhaps $150$\,GeV,
a limit essentially determined by cost, and possibly 
the proximity of the river Rhone.  At $50$\,GeV and $50$\,MW power,
such a configuration may deliver
$5 \cdot 10^{32}$\,cm$^{-2}$s$^{-1}$ luminosity,  
assuming the LHC  was upgraded~\cite{holzer}
to reduce emittances and enhance the proton bunch current.
Because of the constant $e$ current,
unlike the ring situation where the $e$ current
decreases with time, the peak luminosity is worth a factor
of about two more than in the RR scenario.
It varies proportional to $P/E_e$. Integrated luminosities of tens of fb$^{-1}$
appear in reach for the linac-ring configuration.
Efficient use of power may be made if energy recovery (ER)
techniques may be applicable to such a high energy
beam configuration as has been suggested to be worth
seriously considering~\cite{erec}. An ERL linac-ring LHeC had the potential
of  exceeding $10^{33}$\,cm$^{-2}$s$^{-1}$ luminosity,
which, however, requires significant R+D efforts. Energy recovery
as a technique may deserve much higher attention as particle
physics needs to explore smaller and smaller cross sections
while the cost of power and the need for it rise.
\section{Physics}
HERA has taught that an electron-proton collider
at the energy frontier represents a laboratory, in which a wide range of questions can be
studied~\cite{mkry}. 
For the purpose of the CDR,
so far, the attention is focussed to six, related areas of physics as introduced below. 
These are being studied in three working  groups~\cite{uta,olaf,nestor}.   
\subsection{Unfolding the Partonic Structure of the Proton}
The focus of $ep$ is on the structure of the proton.
HERA, despite its success, was limited in kinematic range and luminosity. It therefore
could not exploit fully the potential of $W$ and $Z$ exchange tree level physics
in unfolding the partonic structure of the proton. Its measurements at high $x$ 
are limited, for charged currents (CC) to $x < 0.5$, and at low $x$ to protons only,
and just below where one expects unitarity to limit the growth of the gluon distribution
in DIS.  Much of the present information on parton distributions is only obtained in 
parameterised QCD fits with a number of assumptions on sum rules or
quark-antiquark symmetry. The LHeC, in a kinematic range of $Q^2$ and $x$ much better 
adapted  to the LHC, will be able to measure the valence quarks between a few times
$10^{-4}$ and $0.8$ in $x$, exploiting huge $\gamma Z$ interference effects in
charge and polarisation asymmetries and the abundant 
CC  $e^{\pm}p$ scattering events. The up and down quark distributions
are measured from CC and, using deuterons, employing Gribovs relation 
between shadowing and diffraction to control shadowing  at low $x$.  
Tagging of spectator protons can be expected to essentially remove
Fermi motion effects in $en$ scattering at large $x$.
At the LHeC precision measurements of the charm
and beauty structure functions will be possible over a huge kinematic range, 
around and away from threshold, based on an increased cross section and
on  $c$ and $b$ tagging using dedicated 
tracking detectors and exploiting the fact that the LHeC beam spot will be of
order $10 \times 25$ $\mu$m$^2$, much more narrow than at HERA.
For the first time ever, accurate measurements of the strange and the anti-strange
quark densities  can be performed owing to the high $Q^2$ and large luminosity
in CC $e^{\pm}p$ reactions where charm is tagged. The LHeC further is a single
top and anti-top quark factory with a cross section of order $10$\,pb$^{-1}$
for the $Wb$ fusion cross section in the rather clean single $t$ production environment of
CC scattering. Finally, the much extended range and projected precision
of the measurements, including a dedicated measurement of $F_L$, will
constrain the gluon density over $6$ orders of magnitude in $x$ 
including the edges of the Bjorken $x$ range,
where $xg$ currently is most uncertain. If ever one was interested in the partonic
contents of the proton, here lies an answer which comprises all flavours and
the gluon.
\subsection{Exploration of Superhigh Energy Scales}
The current studies on accessing higher scales than with direct measurements
focus on electroweak couplings, $\alpha_s$ and contact interactions (CI), all exploiting
the precision and kinematic range of simulated LHeC data. The LHeC is an electroweak precision
machine. In initial studies measurements of the light quark weak NC couplings
have been simulated achieving a more than tenfold improvement in accuracy
over current measurements from LEP, the Tevatron and HERA. This hints to a unique
potential for precision electroweak physics reaching high scales and effective couplings.
The experimental accuracy on $\alpha_s$ is of order per mille, ten times better
than currently achieved. This suddenly moves the strong coupling to a level
of accuracy similar to $\alpha$ and $G_F$ when extrapolated to the Planck scale.
Such an accuracy is of crucial importance to explore supersymmetric unification
scenarios. It is also long overdue to precisely measure $\alpha_s$ 
independently of BCDMS data, which
require it to be very small, and thus to possibly resolve a tension between
the DIS and jet based data~\cite{joh}. A tenfold increase in accuracy on $\alpha_s$
would also represent a huge challenge to the techniques and scale dependence
of pQCD calculations.  A study of contact interactions leads to limits of O(50)\,TeV and corresponding results on extra dimensions. The LHeC, owing to its well specified
$eq$ initial state, its kinematic range and 
high precision, is an accelerator for looking into the super-high energy range, i.e.
for much beyond what today is called the Terascale.
\subsection{Complementing the LHC}
The LHC is built as the machine to find new physics beyond the SM by accessing the
Terascale. It relies on the ATLAS and CMS multipurpose detectors and there is a huge
variety of physics predictions to study~\cite{atlas,cms}.  New states may be largely expected
to be produced via gluon ($g$) or $b$-quark induced reactions. An example is
the Higgs particle, $H$. In the SM, $H$ is predominantly produced via $gg$ fusion.
In the MSSM, however, at large $\tan \beta$, a Higgs particle $A$ is produced via
$b \overline{b} \rightarrow A$. The knowledge of the gluon and $b$-quark distributions
becomes crucial particularly at large and at small $x$, which in Drell-Yan scattering
are related to large masses
and rapidities. The best possible knowledge of these and the other parton distributions is
provided by the LHeC, and it should render any pdf related uncertainty in LHC studies 
negligible.
The production cross section of the
Higgs particle at the LHC is not large but the
rarer production modes, as $H \rightarrow \gamma \gamma$
and $H \rightarrow \tau \tau$, have a higher probability to discover and study the
$H$ than the dominating  channel $H \rightarrow b \overline{b}$. That decay, however,
appears to be promising for being studied in $WW \rightarrow H $ fusion in CC
scattering at the LHeC where the unique signature of large missing $E_T$
and efficient $b$ tagging should allow for its decent measurement. 
The cross section is of order $200$\,fb at masses below $150$\,GeV. Therefore
a measurement of the Higgs  coupling to per cent accuracy may be possible.
A special study is being pursued, in which contact interaction
effects, as possibly observed in Drell-Yan scattering at the LHC, are confronted with
the freedom of adjusting parton distributions. In the model studied, accessing $40$\,TeV
CI interference effects, the  parton distributions would be so well constrained
by the LHeC that the CI would be unambiguously identified as new physics, while the HERA
and BCDMS data alone would allow some rearrangement of anti-quark distributions at 
larger $x$ to be compliant with the CI effects and DIS.
The subject of complementing the LHC with $ep$ is being studied further, it
concerns not only genuine signals for physics beyond the SM 
but also subjects like parton dynamics, parton emission and evolution,
factorisation or jet physics, subjects which can be expected to become crucial
for understanding Terascale physics in $pp$ as here it is mainly QCD which leads to BSM.  
\subsection{New Physics in the eq Sector}
An $ep$ machine is naturally suited to search for singly produced new states
coupling to an electron- (or neutrino-) quark pair such
as lepto-quarks or supersymmetric particles in RPV SUSY. The 
cross section for single LQ production is about a hundred times higher than
in $pp$ at a given mass. It depends on the unknown coupling $\lambda$ of the LQ to the electron-quark pair
and means for a coupling  of $O(0.1)$, that LQ masses up 
to nearly $\sqrt{s} = 2\sqrt{E_eE_p}$
may be directly probed. If such states would be discovered at the LHC, most likely in 
pair production~\cite{atlas}, the LHeC appeared most suited to determine
their quantum numbers: the fermion number from a charge asymmetry measurement
reaching higher masses than the LHC~\cite{jinst}, the spin from the angular
distribution of the decay products, possible neutrino decay modes, the coupling down to
small values, and the chiral structure of the coupling using the polarisation of the
lepton beam. At the LHC from LQ pair production and di-lepton production via LQ 
$t$-channel exchange much information may be obtained, a clear spectroscopy,
however, of such states required a polarised, high luminosity $ep$ collider
at the appropriate energies. It is possible that this physics requires a linac to
be chosen for the LHeC as this potentially reaches values of $\sqrt{s}$ of $2$\,TeV
with a $143$\,GeV electron beam or even higher cost permitting. 
A further example for genuine new physics with the
LHeC is the search for excited electrons and neutrinos, which may couple
to a gauge boson via gauge mediated interactions proportional to the
ratio of the unknown coupling to the compositeness scale, $f/\Lambda$,
or phenomenologically similarly to quarks and leptons via
a CI. As has been shown in a first study,
the sensitivity of the LHeC
extends to much smaller values of $f/\Lambda \simeq 10^{-4}$ than the
LHC and possibly to higher masses. Detection of excited fermions would hint
to their compositeness. It is to be noted that the LHC had been chosen for good
reasons, for its high energy and large luminosity, as the pilot machine to 
explore the Terascale. Coming after a decade of LHC experimentation, neither
a lepton-lepton nor an electron-proton collider may be expected to have
a huge resource of
undetected physics to discover at TeV scales. However, their salient features
and complementarity will turn out to be vital for understanding new Terascale physics,
and their salient precision
measurements will lead much further in energy.
\subsection{Parton Saturation at low \boldmath{x}}
At small Bjorken $x$ the gluon and sea quark distributions rise strongly
when $x$ decreases at fixed $Q^2$.  There are arguments based on unitarity
that the gluon distribution should not rise stronger than $Q^2/\alpha_s$GeV$^{-2}$.
Therefore a qualitative change is expected on the parton dynamics at
small $x$, in which non-linear recombination effects are predicted to
modify the so-far successful DGLAP equations. Without a considerable
extension of the kinematic range as compared to HERA,
combined with high precision and large acceptance in forward and 
backward region,  one will not be able to discover parton saturation.
For the CDR on the LHeC, it has been shown that
inclusive structure function measurements, at low $Q^2$ between $0.5$ and
about $1000$\,GeV$^2$ and $x$ down to $10^{-6}$,
of highest possible precision, a per cent on $F_2$ and
about 5\% on $F_L$, will allow to discover saturation in the DIS region.
This would lead to a new understanding of low $x$ theory, as of
resummation and the consistent treatment of $F_L$ and $F_2$. It 
would also be the discovery, in $ep$, of a new
phase of matter, in which partons interact collectively while  $\alpha_s$ is small.
A new regime at low $x$, which long had been searched for at HERA, 
would also directly affect neutrino-astro physics, which deals with 
neutrino-nucleon interactions at extremely small $x$. 
An observation of saturation at the LHeC would perhaps first
be made in diffractive DIS,
in which $xg$ enters squared. With the phase space extended to a few
hundreds of GeV in $M_X$, the diffractive production of charm and beauty states,
of the $Z, W$ and
possibly the Higgs would become a major field of research in $ep$.
From DVCS generalised parton distributions would become measurable
owing to the polarisation and high luminosity. It went without much notice 
that HERA and new theoretical insight has
lead to a significant extension of the parton model with the introduction 
of parton amplitudes and interferences, which are at the basis of proton 
holography~\cite{mueller}. This physics so far is at its infancy.
%
\subsection{Parton Structure of Nuclei}
While any generation of fixed target DIS experiments was given the opportunity
to study lepton-nucleon scattering, i.e. the structure of the neutron and
nuclei besides that of the proton, HERA was not. As a result the knowledge on
the parton distributions in nuclei is restricted to the region of large $x$
and low $Q^2$. It may be extended with the LHeC by four orders of magnitude.
If one wants to understand the plasma effects in $AA$ collisions at the LHC
or to verify theories in nuclear shadowing at low $x$
one needs an adequate experimental  $eA$ input. 
The genuine interest in $eA$ physics is also related to the expectation
that the gluon density in a nucleus $A$ shall be amplified
proportional to $A^{1/3}$. If one assumes $xg$ to grow like $x^{-\lambda}$
towards low $x$, then this amplification implies that a gluon density is
measured at an effective  $x$ value reduced by  $ A^{-1/3\cdot \lambda}$,
i.e. for $x=10^{-4}$, $A=Pb=208$ and $\lambda =0.2$ one would probe
$x$ values close to $10^{-8}$, which is where superhigh energy neutrino
physics is sensitive to.     Such high densities are deep beyond unitarity limits 
and therefore a radically new behaviour of cross sections, $A$ dependences
and parton distributions is expected, with, for example, a logarithmic 
limit $\propto \ln(1/x)$ of the $x$ dependence of $F_2^N$ or a nearly
50\% fraction of diffraction on the inclusive cross section. 
This is in fact is
very high energy physics with extremely large energies $W$ in the proton rest
frame, and particle astrophysics needs to understand the partonic dynamics
and collective behaviour in nuclei much of which may be investigated at the LHeC.
\section{Design}
There is design work ongoing on all relevant components of the LHeC in order
to be able to present a design concept including an evaluation of the
installation and operation interferences with the LHC. It has been assumed in
these studies that the $ep$ collider can operate synchronous to $pp$, i.e.
that it is operational while there still is a significant part of the LHC programme
to be pursued. Extrapolating from the Tevatron or HERA it is likely that the
LHC will be operated for two decades or longer, given the  exceptional
efforts and investments in the machine and  detectors and the expected
broadness of its physics program. Upgrading the LHC with an electron beam
thus appears as a rather natural option, although presently it is impossible
to safely predict the future of the LHC.
\subsection{Accelerator}
The design work on the ring has been divided into ten work packages, the lattice, rf,
injector, beam dump, beam-beam effects, impedance, vacuum, integration and
machine protections, magnets and powering. Similarly there are ten work packages
for the linac, i.e. baseline concepts, rf, positron source, lattice and impedance, beam-beam
effects, vacuum, integration and machine protections, interaction region, magnet
design and powering. There have been  contact persons nominated
to coordinate this work. There have been contributions
made by accelerator physicists from CERN in collaboration with
physicists from BNL, the Cockcroft Institute, Cornell,
DESY, Novosibirsk, Lausanne and SLAC.
With the CDR envisaged for 2010, there is quite some detailed work
ahead and some particular questions deserve special attention. Among those are,
for the ring:
i) the injection using the SPL or alternatives;
ii) the detailed evaluation of a new synchrotron and its installation on top of the
LHC; iii) the design of the bypasses for then existing LHC experiments and
the installation of rf. in these and iv) the availability of crab cavities
to compensate for the small $ep$ crossing angle of $< 2$\,mrad;
and for the linac: i) more detailed designs and selection
of options, as currently one has not decided between pulsed or CW, racetrack or
linear layouts; ii) the evaluation of energy recovery at high energies
and iii) the intensity of the positron source. A common ongoing task
is the calculation of direct and backscattered sychrotron background within
a chosen IR layout, both for the ring and the linac.
 As was mentioned above, from the studies performed so far,
an LHeC, in both options, could be built without considerable further research
and development effort, basically because HERA and LEP have provided
the neccessary experience for a ring and the linac would require TESLA type cavities
of modest, $\sim 25$MV/m, gradient as will be used for the XFEL at DESY.  
In the proposed physics programme, the LHC would have to maintain the
possibility to inject heavy ions, also in the upgraded injector configuration,
and be complemented with a deuteron source and injector.
\subsection{Detector}
The LHeC requires a newly designed $ep$ detector capable of dealing with 
a few TeV of energy in the hadronic and electron final state in forward direction.
Due to the asymmetry in $ep$ beam energies, the low  (high) $x$ programme
requires acceptance of the electron (hadronic final state) close to the beam pipe
in backward (forward) direction. The detector has to be complemented 
by forward taggers for protons, neutrons and deuterons. In backward
direction photons and electrons have to be tagged. The precision demands
are such that the alignment and calibration are assumed to be twice as accurate
as has been achieved with the H1 detector, designed more than 20 years ago. 
Currently a first layout of the inner detector including options for the
solenoidal field exists~\cite{alessandro}.
In this design, the region 
near the beam pipe is occupied with different track detectors, considered 
to be inner pixel layers followed by a GOSSIP (gas on slimmed silicon
pixels) type tracker which combines high resolution
with low material budget. These are surrounded by calorimeter
modules including a Calice type forward and backward
insert for resolving the particle flow. Such a design would 
incorporate the possibility to remove central parts when
focussing magnets need to be placed closer to the IR.
The asymmetry of $ep$ in the current design is reflected in 
different granularity of the forward and backward
detectors, which need to resolve a dense hadronic final
state and to measure backward going electrons, respectively.
A further iteration of the detector will be performed when the
IR calculations of the synchrotron background have converged
as these determine the beam pipe dimensions and thus
the size and layout of the detector to a considerable extent.
The design makes much use of the experience with H1 and
ZEUS, with the LHC detectors and of the considerations for
precision calorimetry for the ILC.
\section{Summary}
The LHeC has so far passed the first obvious tests on its feasibility
and parameters. More will follow, including possibly new ideas and surprises,
and it is for a CDR to comprehensively describe the project and its implications.
The LHeC enjoys increasing attention as has been demonstrated
at the DIS09 workshop, with widening participation or with the decision of NuPECC to now
officially support the LHeC workshop series directed towards the CDR.
The future of particle physics depends on the LHC, its performance
and findings. An $ep$ collider operating at the Terascale has a fundamental
physics programme, a large part of which is independent of these findings.
Only when seen in conjunction with these, however, one understands its
full potential and  the need to reach highest energy and luminosity.
It likely  will be the new physics, which is 
to define the parameters of the $ep$ collider should the CDR be followed by a TDR. 
There
clearly is an exciting future of luminous  deep inelastic scattering, when one thinks
of the exploration of the Terascale and lowest Bjorken $x$ with polarised
$e^{\pm}$ beams attached to the LHC, and including electron-nucleon collisions.  

\vspace{0.2cm}
$\bf{Acknowledgement}$ Many thanks are due to all colleagues engaged with
the LHeC project, 
accelerator, experimental and theoretical physicists, for their efforts to develop the CDR
and contributions which for space reasons have not been cited here individually
but are collected in \cite{lhecweb}. Particular thanks are due to the organisers
for a stimulating DIS09 and their special attention, as with the pre-meeting organisation,
to the LHeC group.

\begin{footnotesize}

\end{footnotesize}
\end{document}